\documentclass{article}

\usepackage{amssymb}

\usepackage{epsfig}

\usepackage{lscape}

\usepackage{graphicx,psfrag,amsmath}

\usepackage{yfonts}

\begin{document}

\def\beq#1\eeq{\begin{equation}#1\end{equation}}
\def\beql#1#2\eeql{\begin{equation}\label{#1}#2\end{equation}}

\def\bea#1\eea{\begin{eqnarray}#1\end{eqnarray}}
\def\beal#1#2\eeal{\begin{eqnarray}\label{#1}#2\end{eqnarray}}

\newcommand{\Z}{{\mathbb Z}}
\newcommand{\N}{{\mathbb N}}
\newcommand{\C}{{\mathbb C}}
\newcommand{\Cs}{{\mathbb C}^{*}}
\newcommand{\R}{{\mathbb R}}
\newcommand{\intT}{\int_{[-\pi,\pi]^2}dt_1dt_2}
\newcommand{\cC}{{\mathcal C}}
\newcommand{\cI}{{\mathcal I}}
\newcommand{\cN}{{\mathcal N}}
\newcommand{\cE}{{\mathcal E}}
\newcommand{\cA}{{\mathcal A}}
\newcommand{\xdT}{\dot{{\bf x}}^T}
\newcommand{\bDe}{{\bf \Delta}}

\def\ket#1{\left| #1\right\rangle }
\def\bra#1{\left\langle #1\right| }
\def\braket#1#2{\left\langle #1\vphantom{#2}
  \right. \kern-2.5pt\left| #2\vphantom{#1}\right\rangle }
\newcommand{\gme}[3]{\bra{#1}#3\ket{#2}}
\newcommand{\ome}[2]{\gme{#1}{#2}{\mathcal{O}}}
\newcommand{\spr}[2]{\braket{#1}{#2}}
\newcommand{\eq}[1]{Eq\,\ref{#1}}
\newcommand{\xp}[1]{e^{#1}}

\newcommand{\tend}[1]{$10^{#1}$}
\newcommand{\tennd}[1]{10^{#1}}
\newcommand{\nrd}[2]{${#1}\times{10^{#2}}$}
\newcommand{\nrnd}[2]{{#1}\times{10^{#2}}}

\def\limfunc#1{\mathop{\rm #1}}
\def\Tr{\limfunc{Tr}}

\def\dr{detector }
\def\drs{detectors }
\def\drsn{detectors}
\def\drn{detector}
\def\dtn{detection }
\def\dtnn{detection}

\def\pho{photon }
\def\phon{photon}
\def\phos{photons }
\def\phosn{photons}
\def\mmt{measurement }
\def\an{amplitude}
\def\a{amplitude }
\def\co{coherence }
\def\con{coherence}

\def\st{state }
\def\stn{state}
\def\sts{states }
\def\stsn{states}

\def\cow{collapse of the wavefunction }
\def\de{decoherence }
\def\den{decoherence}
\def\dm{density matrix }
\def\dmn{density matrix}

\newcommand{\mop}{\cal O }
\newcommand{\dt}{{d\over dt}}
\def\qm{quantum mechanics }
\def\qms{quantum mechanics }
\def\qml{quantum mechanical }

\def\qmn{quantum mechanics}
\def\mmtn{measurement}
\def\pow{preparation of the wavefunction }

\def\me{ L.~Stodolsky }
\def\T{temperature }
\def\Tn{temperature}
\def\t{time }
\def\tn{time}
\def\wfs{wavefunctions }
\def\wf{wavefunction }
\def\wfn{wavefunction} 
\def\wfsn{wavefunctions}
\def\wvp{wavepacket }
\def\pa{probability amplitude } 
\def\sy{system } 
\def\sys{systems }
\def\syn{system} 
\def\sysn{systems} 
\def\ha{hamiltonian }
\def\han{hamiltonian}
\def\rh{$\rho$ }
\def\rhn{$\rho$}
\def\op{$\cal O$ }
\def\opn{$\cal O$}
\def\yy{energy }
\def\yyn{energy}
\def\yys{energies }
\def\yysn{energies}
\def\pz{$\bf P$ }
\def\pzn{$\bf P$}
\def\pl{particle }
\def\pls{particles }
\def\pln{particle}
\def\plsn{particles}

\def\plz{polarization  }
\def\plzs{polarizations }
\def\plzn{polarization}
\def\plzsn{polarizations}

\def\sctg{scattering }
\def\sctgn{scattering}
\def\sctgs{scatterings }
\def\sctgsn{scatterings}

\def\prob{probability }
\def\probn{probability}

\def\om{\omega} 

\def\hf{\tfrac{1}{2}}
\def\hft{\tiny \frac{1}{2}}

\def\zz{neutrino }
\def\zzn{neutrino}
\def\zzs{neutrinos }
\def\zzsn{neutrinos}

\def\zn{neutron }
\def\znn{neutron}
\def\zns{neutrons }
\def\znsn{neutrons}

\def\hf{\tfrac{1}{2}}

\def\csss{cross section }
\def\csssn{cross section}
\def\xrn{x-ray nucleide }
\def\xrnn{x-ray nucleide}
\def\xr{x-ray }
\def\xrs{x-rays }
\def\wv{wavelength }
\def\wvn{wavelength}

\def\intf{interferometry }
\def\intfn{interferometry}
\def\intrf{interferometer }
\def\intrfn{interferometer}

\def\ran{radionuclide }
\def\rann{radionuclide}
\def\rans{radionuclides }

\def\dkm{dark matter }
\def\dkmn{dark matter}

\def\gvtl{gravitational }
\def\gvtln{gravitational}

\def\bh{black hole }
\def\bhn{black hole}
\def\bhs{black holes }
\def\bhsn{black holes}

\def\pbh{primordial black hole }
\def\pbhn{primordial black hole}
\def\pbhs{primordial black holes }
\def\pbhsn{primordial black holes}

\def\dph{$\delta \phi$ }
\def\dphn{$\delta \phi$}
\def\dphg{$\delta \phi_G$}
\def\dphgn{$\delta \phi_G$}

\def\gr{general relativity }
\def\grn{general relativity}

\def\mua{$\mu arcsec$ }
\def\muan{$\mu arcsec$}
\def\rsl{resolution }
\def\rsln{resolution}

\def\lbi{long baseline interferometry }
\def\lbin{long baseline interferometry}

\def\imed{intervening medium }
\def\imedn{intervening medium}

\def\dg{\delta \phi_G}

\title{ Primordial  \bhs in \intf}

\author{ 
L. Stodolsky,\\
Max-Planck-Institut f\"ur Physik
(Werner-Heisenberg-Institut)\\
F\"ohringer Ring 6, 80805 M\"unchen, Germany}

\maketitle

\begin{abstract}
If there is a  population of \bhs distributed 
randomly in space, light rays passing
in their vicinity will acquire random phases. In the ``two-slit''
model  of an interferometer this can, for a high density of \bhsn,
lead  to a diffusion in the  phase difference  between the two arms
of the  interferometer and thus to a
 loss of coherence or ``visibility''in interferometric
observations. Hence the existence of ``fringe constrast'' or
``visibility'' in interferometric
observations can be used to put a limit on the possible presence of \bhs along
the flight path.

 We give a formula for this effect and consider its
application,
particularly for observations in cosmology. Under the assumption
that the \dkm  consists of \pbhsn, we consider sources at high
z, up to the CMB. While the strongest results are for the
CMB as the most remote source, more nearby sources at high $z$
lead to similar effects.

 The effect increases with the baseline, and 
 in the limiting case of the CMB we find that
   with earth-size baselines
 a  non-zero ``visibility'' would
  limit the mass of possible  \pbhsn,   to approximately
$M/M_\odot \leq \tennd{-1}$. Although such limits would not appear
to be as strong as those  obtained, say from microlensing, they involve
a much different methodology and are dominated by very early times (see Table
\ref{tab}).
 Longer baselines lead to more stringent limits and
in principle with extreme lengths,
 the method could possibly  find positive evidence for \pbhsn.
In this case, however, all other kinds of phase averaging would
have
to be constrained or eliminated. 
\end{abstract}

\section{Introduction}

When  there are  \sctg centers between a source of radiaton and
an observer, the associated  deflections 
  lead to a spread of the incoming directions.
The usual concern with this effect 
is  the  intrinsic  limitation on the     angular \rsl  it presents
 for the observer.
  But if the nature of the \imed or \sctg centers is unknown,  it
can also be used to obtain information on  these.

 In a recent paper\cite{sra} we discussed  how  this 
could  be used to limit, or perhaps to  discover 
\pbh \dkmn.
With  the CMB as the source \cite{joe}  and assuming  the 
\dkm  composed of \bhsn, we found that there is a limit 
on the angular \rsl $\delta \alpha_{lim}\approx 
\nrnd{3}{-7}\sqrt{M/M_\odot}~rad$. Combining this with
 the  milliradian \rsl demonstrated \cite{pl}
in the observations of  ``acoustic peaks ''
one concludes that the  (average) mass $M$  of possible
\bhs  is constrained to 
 about $M/M_\odot\leq \tennd{7}$.

  The \sctg process in question is  the \gvtl
deflection of \phosn, as in the bending of light by the sun, 
\beql{bndg}
\delta \alpha=\frac{4GM}{b}= \frac{2r_s}{b} 
=\nrnd{6.3}{-13}( ly /{b})\, ( M/M_\odot)  \,,
\eeql
with $\delta \alpha$ the deflection angle in radians, $b$ the
impact parameter of
the ray or \phon, and $r_s=2GM$ the schwarzchild radius associated
with
 $M$,  the mass of the \bh or gravitating object
 (c=1 units, ly=light year, $M_\odot$=solar mass).

At present the most likely explanation for the \dkm seems to be 
a new kind of elementary \pln, which might  be detected in the
laboratory.
 especially by 
the cryogenic technique \cite{cryo}, But \bhs are an interesting
alternative
\cite{carr}. 
It would evidently be interesting  to be able to extend our
argument to smaller
angles, and so to eliminate, or conceivably to discover,
 smaller mass \pbhsn.

 However, there appears to be a difficulty.
  There are very
likely  no  very fine, prominent features on the CMB, analogous to
the ``acoustic peaks'', but subtending very small angles. How then,
could one
 exhibit a   high \rsln, or the lack of it, when there are no fine
feaures to observe?

\section{Principle }

Here we would like to propose an solution to  this question, 
 and to apply it to the possible  \pbhs at the CMB.
  As briefly alluded to in \cite{sra},  the answer  lies in
examining the coherence of the wavefront at
large separations. If instead of a perfect plane wave from some
source,
one has  an incoherent sum of slightly differing incoming
directions, 
the phases at widely separated points will become uncorrelated or
incoherent. That is,  the coherence over the wavefront at large
separations is closely
connected to the angular spread, the limiting \rsln, of  the
incoming waves.  

  In \lbin, which is used in astronomy to achieve high angular
\rsln,  one has a form of the two-slit
interference arrangement \cite{haj}. The  high \rsl arises from  a
long baseline
$B$ between   two ``slits'' or receivers so that
$\delta \alpha_{min} \approx \frac{1}{2\pi}\frac{\lambda}{B},$
 with $\lambda$ the wavelength of the radiation. 
 For \lbi to function it is evidently necessary that the two
paths from the source to the ``slits'' be coherent. If the two rays
which
should ultimately interfere encounter different perturbations or
\sctgn s along the way to the receivers, there will be a loss of
interference and
a breakdown of the method, giving an instrinsic limiting \rsln.

When  $B$ is small and the two paths traverse essentially the same
environments they will be coherent. As $B$ increases,  at some
point this will no longer be true and interference breaks down.
 In the following we examine how this would happen due to 
 the presence of \bhs along the flight path from the CMB, or possible
more nearby sources.
 The resulting  loss of ``visibility'' or ``fringes'', or the
absence of such loss,  can
then be used as a limit, or  an indication for the \bhsn.

\section{Calculation}

The basis for the calculation will be the formula for the phase
acquired by a \pho  or light ray  traversing a weak gravitational
field. 
In addition to the plane wave term in the phase, their is an
 extra \gvtl phase $ \phi_G$ which 
 according to general arguments \cite{ml} is
\beql{gphi}
    \phi_G=\hf\int h_{\mu\nu}k^\mu dx^\nu
\eeql
 where $h_{\mu\nu}$ is a  small  deviation from an overall  metric
tensor,  $k$ the four-momentum or wave vector of the \phon, 
 and the integration is along the path of the ray.

We   need  the difference in phases
$\delta \phi_G$ for two parallel rays 1,2  in the vicinity of a
\bhn,
with impact parameters $b_1,b_2$. 
Integrating  the difference according
 to  \eq{gphi}  
 with  the schwarzschild metric  and 
 taking
the difference  $\delta b=b_1-b_2$ to be small compared to that for
the
 pair as a whole, $b$,  
 one finds,

\beql{hterm2}
 \delta \phi_G= \omega r_s \delta b \int_{-\infty}^\infty  dz\, 
\frac{d}{d b}  
\frac{1}{\sqrt{z^2 +b^2}}=2 \omega r_s \frac{\delta b}{b} \, ,
\eeql
where $r_s$ is the  schwarzschild radius of the body, and
$\omega=k$
the frequency of the radiation.
 The phase difference
increases linearly with the ray separation $\delta b$.
The assumption $\delta b/b <<1$ seems justified since while
 $\delta b$ will be given by the baseline, and so be on the of
order of 
 earth or perhaps solar system size,
we anticpate $b$ on the order of light years  or more.

There is a close relation between the classic bending formula
\eq{bndg}
and \eq{hterm2}.  When $b_1$ and $b_2$  lie along a radius vector,
 this calculation for  $\delta \phi_G$ is the same as needed for
the bending of a wavefront, where the \sctg angle is given by the
transverse
derivative of the phase. That is,  for 
a  \sctg process with long range forces and smoothly varying phase 
 $\phi(b)$,  one has for the bending of a plane wave
$\delta \alpha= k_T/k=\frac{1}{\omega} \frac{d \phi(b)}{db},$
where $k_T$ is the transverse momentum to the \phon. 
In  \eq{hterm2} this amounts to dividing by $\omega\, \delta b $ 
which  yields, in fact, \eq{bndg}.

\section{Random  $\delta \phi_G$ }
A pair of rays starting at a point on the source and arriving at
the
two `slits' or receivers will experience different fields and so
acquire
different phases along their  paths.
 If the \dkm is made of \bhsn, presumed random in their
locations, this will lead to a random $\dg$, which. if it is large
enough,
will lead to loss of coherence or
``visibility'' in the interference pattern.

In view of the small values of the factors in \eq{hterm2},
  $\delta \phi_G$
from a single passage near a \bh will be small. However with many
repeated passages, which would occur if the \dkm is made of \bhsn,
it
is possible that a substantial $\delta \phi_G$ will  acccumulate. 

The sign  of $\delta \phi_G$ in a single passage
  depends on the orientation
of the ray pair with respect to  the radius vector from the \bhn,
sometimes
ray 1 will be closer to the \bh and sometimes ray 2, so that the 
average value of $\dg$ will be zero.
 However  the
accumulated $\delta \phi_G$ will fluctuate, with the character of
a random
walk.  In such problems the expected magnitude of the quantity, 
is characterized by the variance, here $\overline{(\delta
\phi_G)^2} $
 and it is this quantity that we wish to study.

When the two rays are combined in the \drn, the $\dg$ will lead to
fluctuating factors such as $cos (\dg)$ multiplying the
interference term.
The significance of $\overline{(\delta \phi_G)^2}$ may be seen with
a
gaussian probability  distribution $P_{gauss}$ \cite{pdg}  for
$\dg$:   
\beql{gau}
\overline{ cos(\dg)}=\int d(\dg)\, cos(\dg)\,  P_{gauss}=e^{-\hf
\overline{(\delta \phi_G)^2}}
\eeql
When  $\overline{(\delta \phi_G)^2}$  becomes larger than one,
interferences are ``washed out''. We shall use this condition,
\beql{cndtn}
\overline{(\delta \phi_G)^2} > 1
\eeql
to estimate when coherence is lost and
 the effects of the possible \bhs would  become significant. 

\section{Calculation of $\overline{(\delta \phi_G)^2}$}
To calculate the total  variance  we 
 take the individaul \bhs to be uncorrelated
 and use the principle of the addition of variances.
We do this by calculating the variance from one  encounter 
 as given by \eq{hterm2} and then
multiply by the number of encounters along the flight path.

These encounters will take place in the general background
spacetime, which
we take to be that of the FRW metric given by  $ds^2=dt^2-a^2d{\bf
r}^2$ 
 with expansion factor $a=(t/t_{now})^{2/3}$,
$t_{now}=\nrnd{2.9}{17}s$
\cite{pdg}.
In this connection 
 one notes that \eq{gphi} is a scalar in \grn, being the
contraction 
of two tensors. Thus we may use the  evaluation \eq{hterm2},
 carried out for an ambient flat space,  in the local frames of   
\bhs 
  along the flight path and simply add the various $(\dg)^2$.  Of
course the parameters needed, such as the frequency, and
 the separation of the two rays, will
 depend on the location in the general background spacetime.

There is a difficulty, however, 
 associated with the long ranged
nature of the gravitational field. In order to define the
probability
of an ``encounter'' one must assign some range or cross section to
the
field around each  constituent, here the \bhsn. But for the
gravitational field
there is no such parameter, the force has an infinite range. 
We shall handle this problem by assuming that the picture of a
single,
independent constituents
 holds up to some distance $b_{max}$, with a corresponding \csss 
$\sigma =\pi b_{max}^2$.   
  This seems a reasonable procedure
since the contributions to $\overline{(\delta \phi_G)^2}$
 are largely at small $b$.
The  $b_{max}$ parameter can be plausibly be taken
to be the typical distance between \bhsn, 
 We shall see below that the choice of  $b_{max}$ enters into the 
final
result only logarithmically.

\subsection{Number of Encounters}
With this understanding, we first consider the number of \bhs
encountered
in an interval of cosmic time $dt$.
 Along the flight path, the
 number of \bhs per unit area in a time interval $dt$ is
  $d\rho_2=\rho(t) \,dt$, where $\rho(t)$ is the number density.
Associating with each \bh a  ``\csss'' $\sigma$,
one has for the number of encounters $dN$ in a time $dt$

\beql{dn}
dN= \sigma d\rho_2=\rho(t)\, \sigma \, dt
\eeql

\subsection{Single Passage}
For the $\overline{(\dg^2)}_{single}$ associated with  a single
encounter,
we  assume the $b$ for the ray pair to be uniformly distributed
across the
area of $\sigma$,  integrate the square of \eq{hterm2} over this
area and divide by the area.

{\bf Angular Factor:}
First  we consider
 the angular factor involved in $\delta b$, which is the origin of
the
fluctuating sign.
 Let $\bf b_1$ and $\bf b_2$  be the respective vectors from the
\bh  to the
 points of closest approach, and
$\bf d=(b_1-b_2)$  the vector connecting them. Then 
for the $\delta b$ needed in \eq{hterm2} one finds
\beql{ort}
 \delta b= {b_1-b_2}= 2{\bf d}\cdot \frac{\bf b}{2b}=d\,cos\theta
\,,
\eeql
with $\theta$ the angle between $\bf d$ and the radius vector
 $\bf b=\hf(b_1+b_2) $. This relation follows
from writing $\bf b_1,b_2=  b\pm \hft d$, squaring, taking the
difference, and using $b_1+b_2\approx
2b$. For the nearby universe and the  line of sight perpendicular
to
the baseline, $d$ is
simply  the baseline $B$.(In the following we take the direction of
observation to be a right angles to the baseline, otherwise a
trignometric
factor can be included. )
For the angular integration one thus obtains
\beql{cosav}
\int_0^{2\pi}d\theta (\delta b)^2=\pi d^2
\eeql

{\bf Integration Over b:}
For the remaining $b$ integration one now has
\beql{single}
\overline{(\dg^2)}_{single}=\frac{1}{\sigma }
  4\pi (\omega r_s d)^2 
 \int_{b_{min}}^{b_{max}}\frac{db}{b}=
\frac{1}{\sigma}4\pi (\omega r_s d)^2 
ln(\frac{b_{max}}{b_{min}})\,,
\eeql
with  $b_{min}, b_{max}$  some smallest, largest, impact
parameters.

\subsection{General Formula}
\eq{single} must now be multiplied by the number of encounters
in an interval of cosmic time.  
\beal{difvar}
d\overline{(\dg^2)}&=& dN\times \overline{(\dg^2)}_{single}\\
\nonumber
&=& dt \rho \, \sigma 
\frac{1}{\sigma}4\pi (\omega r_s d)^2 
ln(\frac{b_{max}}{b_{min}})\\
&=&dt \, 4\pi  \rho \,\,
  (\omega r_s d)^2
  ln\bigl(\frac{b_{max}}{b_{min}}\bigr)\,.
\eeal
As anticipated,  $\sigma $ has cancelled and the  $b$ 
limits are only in the logarithm. The appearance of the logarithm
is an
interesting feature of the present problem, in contrast to the more
familiar
situation of \pls or \phos passing through ordinary neutral
 matter. There one deals with the shielded coulomb field or the
short
ranged nuclear force, while here one has the unshielded, long
range, \gvtl
force. 

It remains to insert values for $b_{max}, b_{min}$. For $b_{max}$
we take
the distance between the \bhsn, namely $\rho^{-1/3}$.  Beyond this
 distance one must
consider the summed fields from many objects, leading to smooth and
relatively
weak fields. For the lower limit  we take  $b_{min}=r_s$.

The final result is thus
\beal{fnl}
\overline{(\dg^2)}
&=&4\pi\, r_s^2\int dt \,  \rho(t)
 \biggl (\omega(t)  d(t)\biggr)^2
  ln\bigl(\frac{\rho^{-1/3}(t)}{r_s}\bigr)\\
\nonumber
&=&16\pi^3\, r_s^2\int dt \,  \rho(t)
\biggl (d(t)/\lambda(t)\biggr)^2
  ln\bigl(\frac{\rho^{-1/3}(t)}{r_s}\bigr)
\eeal
writing $\omega d=2\pi d/\lambda$, and
where we have explicitly indicated those quantities which might
depend
on $t$.

{\bf Parameters:}
For  quantitative estimates, we shall use light years (ly) for
lengths
and solar masses ($M_\odot$) for mass. Thus the number density is
 $\rho(t)=\rho_o(t)(M_\odot/M)/ly^3) $ for a \dkm mass density
$\rho_o(t) M_\odot/ly^3$ ,
where $\rho_o(t) $ is a dimensionless density  parameter. We will
be
assuming that a well defined mass or average mass can be assigned
to the
\bhsn, if this is not true the arguments may have to be
reconsidered.
Furthermore, anticipating the use of earth-size baselines and
microwave
wavelengths, as in ref.\cite{eht},  we introduce a length parameter
$d_o=\nrnd{1.0}{4}\, km$ and a
wavelength parameter $\lambda_o=\nrnd{1.0}{-1} cm$.
 Collecting all factors and
 inserting  $r_s=3.0\, km(M/M_\odot)= \nrnd{3.2}{-13}(M/M_\odot)ly$
as  in  \eq{bndg},
\eq{fnl} becomes
\beal{fnl1}
&\overline{(\dg^2)}&
=(\nrnd{5.1}{-3})(M/M_\odot)\int dt/y \,  \rho_o(t)
\biggl(\frac{ d/d_o}{\lambda/\lambda_o}\biggr)^2
  ln\bigl(\frac{\rho^{-1/3}(t)}{r_s}\bigr)\\
\nonumber
&=&(\nrnd{5.1}{-3})(M/M_\odot)\int dt/y \,  \rho_o(t)
\biggl(\frac{ d/d_o}{\lambda/\lambda_o}\biggr)^2
\biggl((1/3)ln\rho_o(t)-(2/3) ln(M/M_\odot)+ 29)\biggr)\\
\nonumber
&=&(\nrnd{1.5}{-1})(M/M_\odot)\int dt/y \,  \rho_o(t)
\biggl(\frac{ d/d_o}{\lambda/\lambda_o}\biggr)^2
\biggl(1+ (0.011)ln\rho_o(t)-(0.023)ln(M/M_\odot)\biggr)
\eeal
( Since we have $c=1$ units and are dealing with
light rays, one may use a distance label $r$ instead of $t$. In
that case
one writes $\int dt/y\to \int dr/ly$). We also note that with this
notation
the dimensionless quantity  $\int dt/y\,\,\rho_o $ is simply the
coefficient in 
the mass column
density; that is, the  mass column density is $\int dt/y\,\times \rho_o
\times 
M_\odot/ly^2 $.

Because of the small coefficient of the $ln\,\rho_o(t)$ term, it
may be
possible to replace it by some typical value and move it outside
the
integral. In that case 
\eq{fnl1} becomes
\beal{fnl2}
\overline{(\dg^2)}
&\approx&C\,\int dt/y \,  \rho_o(t)
\biggl(\frac{ d(t)/d_o}{\lambda(t)/\lambda_o}\biggr)^2\\
\nonumber
C&=&(\nrnd{1.5}{-1})(M/M_\odot)
\biggl(1+ (0.011)ln\rho^{typ}_o-(0.023)ln(M/M_\odot)\biggr)
\eeal
As should be expected,
 the effect vanishes with $M\to 0$, as the mediun becomes smooth
and less ``lumpy''.

 In these estimates we have assumed
a uniform and constant relation between the \bhs and \dkmn. Variations
on this assumption can be taken into account by changes in the
density factor $\rho$ in the above.  If for example, one would like to
assume that only a certain fraction of the \dkm is \bhsn, the fact that
that $\overline{(\dg^2)}$ is linear in the density implies that the results
need only to be reduced by this fraction.  In addition, our assumption
that the \bhs are distributed randomly could be violated if there is
a  tendency for them  to ``clump''. Such concentrations, however, would
amount to a small  increase
in the effective  mass $M$ in our formulas, for the
 given amount of \dkmn, and  so would only increase the
effect. This originates from the factor $ r_s^2\sim M^2$ in the formulas.

\section{CMB}

We now consider the application of the formula to rays traveling
 from the CMB to the Milky Way.
For the density we take 
\beql{r0}
\rho_0\approx\frac{1}{a^3(t)}(\nrnd{1.0}{-9})\,,
\eeql
as follows from  taking  the present cosmological  \dkm \yy density
  at 1/4 
the critical density  \cite{pdg}, and  applying the cosmological
expansion
factor $a(t)=(t/t_{now})^{2/3}$.

To simplify the integral  we  replace the time dependent
 $ln\,a(t)$ arising in $ln\,\rho_o(t)$ 
 by its value  at $t_{cmb}$, namely $-7.0$. Since  $ln\,a(t)$
varies from zero to this value in the time integration, this 
simplification
 gives  a small  underestimate of the total integral. This gives
$ln\,\rho^{typ}=0.3$ and  one then has for $C$ in \eq{fnl2}
 $C=(\nrnd{1.5}{-1})(M/M_\odot)
\bigl(1 -(0.023)ln(M/M_\odot)\bigr)$ and so
\beql{fnl3}
\overline{(\dg^2)}\approx \nrnd{1.5}{-10}(M/M_\odot)
\bigl(1 -(0.023)ln(M/M_\odot)\bigr)
\int dt/y \,  \frac{1}{a^3(t)}
\biggl(\frac{ d(t)/d_o}{\lambda(t)/\lambda_o}\biggr)^2
\eeql
Since $\int dt$ will be at least \tend{10} years,
$\overline{(\dg^2)}$
can easily be greater than one, and satisfy the condition
\eq{cndtn} .

\subsection{Time Integration}
We are thus left with the integral
\beql{difvar34}
 I =  \int_{t_{cmb}}^{t_{now}} dt/y\,  \frac{1}
{a^3(t)}  \biggl(\frac{ d(t)/d_o}{\lambda(t)/\lambda_o}\biggr)^2 
\eeql
to evaluate.

{\bf Frequency Shift:} In the rest frame of a \bh at time $t$, one
has
 the frequency
 or wavelength  red shifted, so that if $\omega$
 is the frequency of  observation,  one  has
$\omega \to \omega/a(t)$ or $\lambda \to  a(t)\lambda$.

{\bf Ray Separation:} \label{sep}
The remaining question is the ray separation $d(t)$. The
rays must originate from a common point to interfere. 
 In euclidean space, without general relativistic  considerations,
 one would have
 $d(t)=\frac{B}{t_{now}-t_{cmb}}(t-t_{cmb})$, where we chose
 the constants to give $d=0$ at emission and $B$ at the baseline
of the \drn.

 To transfer this reasoning to the general relativistic
 situation, we use the ``conformal
time" $\eta(t)=3t_{now} x^{1/3}$, and 
 introduce the dimensionless  time variable $x$ so that 
\beql{x}
x=t/t_{now} ~~~~~a=x^{2/3}~~~~~\eta=3 \,t_{now} x^{1/3}
~~~~~x_{cmb}=\nrnd{2.7}{-5}~~~~~
\eeql
In terms of $\eta$ one has
 $ds^2=a^2(d\eta^2-d{\bf r}^2)$. Light rays travel
in ``straight lines'' in these coordinates, so   for the coordinate
separation, one has the euclidean result as before,
$d_{coord}(\eta)=\frac{B}{\eta_{now}
-\eta_{cmb}}(\eta-\eta_{cmb}).$ 
 However, we require the physical separation so $a(t)d_{coord}(t)
$ should
 be used for $d$.
 The $a$ factors  cancel then in the ratio $d/\lambda$ 
and the integral becomes

\beal{difvar5}
I& = & (t_{now}/y)  \biggl(\frac{B/d_o}{\lambda/\lambda_o}\biggr)^2
  \int_{x_{cmb}}^1 dx\,  \frac{1}
{x^2}\biggl(\frac{\eta-\eta_{cmb}}{\eta_{now}-\eta_{cmb}}\biggr)^
2\\
\nonumber
& = &(t_{now}/y)  \biggl(\frac{B/d_o}{\lambda/\lambda_o}\biggr)^2
  \int_{x_{cmb}}^1 dx\,  \frac{1}
{x^2}\biggl(\frac{x^{1/3}-x^{1/3}_{cmb}}{1-x^{1/3}_{cmb}}\biggr)^2
\eeal

In \eq{difvar5} we have a convenient separation into factors
depending on
the observational arrangenment ($B,\lambda $) and  a ``cosmological
 integral''  $\int_{x_{cmb}}^1 ...$ independent of the setup. 

\subsection{CMB Result}
Examination of the integrand shows that with
$x^{1/3}_{cmb}=\nrnd{3.0}{-2}$
 from \eq{x} , it is strongly peaked around
$x\sim \tennd{-4}$. Since $x_{cmb}=\nrnd{2.7}{-5}$, this shows that
the
effect would come predominantely close to the CMB and that a
positive observation would support the idea that the \bhs are
primordial. For
the integration one  finds the integrated value 
  $\int_{x_{cmb}}^1 ...= 32 $.
As in \cite{sra}, there is an enhancement at early times,    
but  due to the requirement of the convergence of the rays at their
origin.
to a much lower power.
Combining with \eq{fnl3} one has finally,

\beal{intapp}
\overline{(\dg^2)}
&=& (\nrnd{4.5}{-9})\bigl(M/M_\odot\bigr)  (t_{now}/y) 
\biggl(\frac{B/d_o}{\lambda/\lambda_o}\biggr)^2
  \\
\nonumber
&=&  (\nrnd{4.1}{1})\bigl(M/M_\odot\bigr)
  \biggl(\frac{B/d_o}{\lambda/\lambda_o}\biggr)^2\biggl( 1
-(\nrnd{2.3}{-2})\,ln(M/M_\odot)\biggr)\,,
\eeal
after inserting $t_{now}=\nrnd{2.9}{17}sec=\nrnd{0.92}{10} y$.

\subsection{ Sources More Nearby }\label{nearer}

We have chosen to evaluate the  effect with the CMB as source
since this is
at the greatest distance available. However interferometry on the
CMB \cite{dasi} has, until now,  not been with the earth-sized
baselines as used by the EHT project \cite{eht}. 

 It may thus be of interest to inquire as to the size of the
effect for  more nearby sources, say for $z\leq 10$, where there
can be
localized sources. To do this, we must
evaluate the integral $ \int_{x_{cmb}}^1... $ in \eq{difvar5} with
$x_{cmb}$ replaced  by the $x$ of the source, which we shall call
$x_o$.

In the Table we show the results for a few values of $x_o$. The
first
two columns
gives the `distance' of the source expressed as $x_o$ or
equivalently
the redshift factor  $z_o$.
The second column gives the location of the peak of the integrand,
 in terms of its  $x$ or  $z$. Finally, the
last column shows the value of the integral $ \int_{x_o}^1... $.
For
comparison  the last row shows the CMB result from the subsetion
above.
The integral is given by the expression  $ \int_{x_o}^1... =
(1/1-x_0^{1/3})^2(x_0^{-1/3}-3+3\,x_0^{1/3}-x_o^{2/3}).$

\begin{table}
\begin{center}
\begin{tabular}{|l|l|l|l|}
\hline
``Distance''$x_o$ &"Distance" $z_o $ &Peak Location\,$x,
z$&$\int_{x_o}^1...$\\
\hline
\hline
$0.032$&$8.9$&$0.1,~~~~3.6$&
$1.4$\\
\hline
0.08&4.4&0.25~~~~1.5 &1.3\\*
\hline
$0.10$&$3.6 $ &$0.35~~~~1.0$&
$1.2$\\
\hline
$\nrnd{2.7}{-5}$&$\nrnd{1.1}{3}$&$\tennd{-4}~~~~460$&32 \\
\hline
\end{tabular}
\end{center}
\caption{Values of the ``cosmological integral' $\int_{x_o}^1...$ in
\eq{difvar5} for some sources in the range $1<z<10$, and finally in the last
row,  for the CMB. The first column gives the scaled time
variable $x_o$ for the source, the second column its redshift factor $z_o$
and the third column where the integral has its peak in terms of these
quantities. One notes that while the effect is largest for the CMB,
 by about a factor 20, the other estimates are not enormously smaller. }
\label{tab}
\end{table}

One sees that the effect for such nearer sources would be reduced
by a factor
of only about twenty or thirty compared to  the CMB,
 despite the very great differences in ``distance''.

\section{Galactic Contributions}\label{gal}
On their  way to the earth, rays from the CMB or other sources will
also pass
through the
galaxy, first presumably through  a large, extended, but tenuous
dark halo, and then through some part of the more dense inner
galaxy.
Can these
traversals make a significant contribution to the above estimates?

A thorough answer to this question would  require the use of
detailed models
of the \dkm distribution in the Milky Way and its halo, and refered
to a
particular line-of-sight. However a crude estimate suggests the
answer is negative, at least for the CMB as source.

First for the inner galaxy,  the estimated
 \dkm density in the vicinity
of the earth \cite{pdg} is $\rho_o=\nrnd{3.0}{-4}.$
 (Units of $M_\odot /ly^3$).
 Taking a typical path length of \tend{4}ly at this density, on
finds from \eq{fnl2}
that the numerical coefficient in front of \eq{intapp} becomes 
\nrd{4.5}{-1} instead of \nrd{4.1}{1}, about two orders smaller. 

For the dark halo, the density presumably drops off quickly beyond
some
radius $R_{inner}$. If we crudely take  the density beyond this to
vary as $\sim 1/R^2,$
as one would have with a constant rotation curve, then 
for the column depth integration through the halo one has 
 $\rho_{inner} \int_{R_{inner}}^{\infty}dR
(R/R_{inner})^{-2}=\rho_{inner} R_{inner},$ qualitatively the same
or perhaps
somewhat larger than that for the inner galaxy.

 It thus appears that the
galactic contributions to $\overline{(\dg^2)}$ are one or two
orders less
than  the cosmological \eq{intapp}. In the event of a possible
positive
signal from more nearby soures, as in subsection \ref{nearer}, it
might
be necessary to disentangle these from the local, galactic ones.

  These argument concerning the
galactic
contribution  are of course quite
schematic, necessitating  more detailed estimates.
 In this connection, we have
recently \cite{joe}  estimated the limits arising from models of
 the `spike' of \dkm
believed to  surround 
the supermassive \bh in M87, using the arguments of
ref.\,\cite{sra}, 
and the angular \rsl of the EHT results \cite{eht}.

\section{Conclusions}

The examination  of interferometric  ``visibility'' at long baselines provides
a very high sensitivity method for studying an \imedn. We have
appplied the method to the question of \pbhs in cosmology.

Interferometric observations on the CMB have been considered
and  carried out over the
last decades  \cite{dasi}. If observations
 with earth-size baselines, as in
\cite{eht},
reveal coherence, a nonzero ``visibility'',  then  according to
\eq{intapp} 
and \eq{cndtn}, \pbhs down to the
 $(M/M_\odot)\sim \tennd{-1}$  range are ruled out.
Longer baselines would of course  access  smaller $M$. An
earth-moon
baseline, $\sim \nrnd{4}{5}\, km$, for example, leads
  to  $(M/M_\odot)$ in the $\sim \tennd{-4}$ range.

 These values are similar but generally weaker than  limits  found
in work of the ``macho search '' or
 ``microlensing'' type 
\cite{machos}. An overview of the  limits from different
methodologies is  provided
by a plot in reference \cite{joe}. But the interferometric method
  would provide  a  different  approach to the question,
and also refer to quite different epochs in cosmology.
The microlensing work uses background
 stars in the nearby universe, Andromeda or the Milky Way bulge,
while the main contribution to  \eq{intapp}, as said and as is shown
in the Table, is at very high
redshift, close to the CMB. 

In subsection\,\ref{nearer} we have discussed applying the method
to
 sources in the
range $1<z<10$, leading to somewhat smaller effects. This region
could involve different technologies and would cover a different 
region in cosmology.

\section{Remarks}
Conceivably, the method could  also be   used to find 
   positive evidence for \pbhn s.
 This  would be
evidently a more difficult matter than the simple setting of
limits,
since all kinds of other phase averaging,
 instrumental as well as natural, would have to be eliminated
or controlled. In this
respect, the simple frequency dependence of \eq{intapp}, which
originates in the
achromatic behavior of \phos  in the \gvtl field, might be helpful.

Our discussion has been for the
 `two slit' configuration, which is
the paradigm of astronomical \intf \cite{haj}, while  contemporary
\lbi  often uses    many baselines together.
   But our
 essential effect, due to  interfering rays traversing 
different fields,  should be dominated by the largest baseline.
For the more complex , multi-baseline
arrangements it would be interesting to examine possible
improvements in sensitivity and statistics, but 
 the ``two slit'' estimate  for the sensitivity to
potential \bhs 
 should remain roughly correct,

We stress that our considerations,
 or equivalently those  concerning  the   \rsln, should not be
confused 
with the question of the power spectrum of the CMB. 
For the present purposes the role of the CMB is simply to
provide  a remote ``source'' and we are not concerned with the
nature of the CMB itself. This point is discussed  in section 5.5
of
\cite{sra}.

\end{document}